# Hydex Glass: a New CMOS Compatible Platform for All-Optical Photonic Chips

D. J. Moss[1,2*], S.D.Jackson[1**], A. Pasquazi[2], M. Peccianti[2,3] and R. Morandotti[2]

[(1)] School of Physics, University of Sydney, New South Wales 2006 Australia
[(2)] INRS-EMT, 1650 Boulevard Lionel Boulet, Varennes, Québec, Canada, J3X 1S2
[(3)] IPCF-CNR, UOS Roma, "Sapienza" University, P. A. Moro 2, 00185 Rome, Italy
*Current address: School of Electrical and Computer Engineering, RMIT University, Melbourne, Vic., Australia 3001.
**Current address: School of Electrical Engineering, Macquarie University, Sydney, NSW, Australia.
david.moss@rmit.edu.au

*Abstract*
We demonstrate a range of novel functions based on a high index doped silica glass CMOS compatible platform. This platform has promise for telecommunications and on-chip WDM optical interconnects for computing.

## I. Introduction

All-optical signal processing has been demonstrated extensively in silicon nanowires [1] and chalcogenide glass (ChG) waveguides [2] including demultiplexing from 160Gb/s [3] to over 1Tb/s [4] via four-wave mixing (FWM), optical regeneration [5,6] and many other processes. The third order nonlinear efficiency of all-optical devices can be improved dramatically by increasing the waveguide nonlinear parameter, $\gamma = \omega\, n_2 / c\, A_{eff}$ (where $A_{eff}$ is the waveguide effective area, $n_2$ the Kerr nonlinearity, and $\omega$ the pump frequency) as well as by using resonant structures to enhance the local field intensity. High index materials, such as semiconductors and ChG, offer excellent optical confinement and high values of $n_2$, a powerful combination that has produced extremely high values of $\gamma$: 200,000 $W^{-1}\,km^{-1}$ for silicon nanowires [1], and 93,400 $W^{-1}\,km^{-1}$ in ChG nanotapers [2]. Yet silicon suffers from high nonlinear losses due to two-photon absorption (TPA) and the resulting free carriers. Even if free carriers can be eliminated by using p-i-n junctions, its poor intrinsic nonlinear figure of merit (FOM = $n_2 / (\beta\, \lambda)$, where $\beta$ is the two-photon absorption coefficient) is very low. While this FOM is considerably higher for ChG, the development of fabrication processes for these newer materials is at a much earlier stage. The potential impact of a low FOM was dramatically illustrated in recent experiments in silicon at longer wavelengths below the TPA threshold [7, 8]. While TPA can, in some instances, be turned around and used to advantage for all-optical functions [9, 10], for the most part in the telecom band the low FOM of c-Si poses a fundamental limitation and is a material property that cannot be improved.

We report a range of novel lasers based on a CMOS compatible high index doped silica glass platform [11- 18] that exhibits virtually no nonlinear absorption in the telecom band. We demonstrate an integrated multiple wavelength source [13], an ultra-high repetition rate modelocked laser [15], a dual comb oscillator [17], and a stable self-locked OPO [18] based on integrated ring resonators. We achieve CW optical "hyper-parametric" oscillation in a high index, high Q factor (1.2 million) silica glass micro-ring resonator with a differential slope efficiency above threshold of 7.4% for a single oscillating mode out of a single port, a CW threshold power as low as 50mW, and a controllable range of frequency spacing from 200GHz to more than 6THz. We also achieve modelocking in this device with pulse repetition rates as high as 800GHz [15]. Finally, we report novel functions in ultra-long (45cm) spiral waveguides, including a device capable of measuring both the amplitude and phase of ultrafast optical pulses, using an approach based on Spectral Phase Interferometry by Direct Electric-field Reconstruction (SPIDER) [16]. The success of these devices is due to their very low linear loss, a high nonlinearity parameter of $\gamma \cong 233\,W^{-1}km^{-1}$ as well as negligible nonlinear losses up to extremely high intensities (25GW/cm$^2$) [14]. The low loss, design flexibility, and CMOS compatibility of these devices will enable multiple wavelength sources for telecommunications, computing, metrology and other areas.

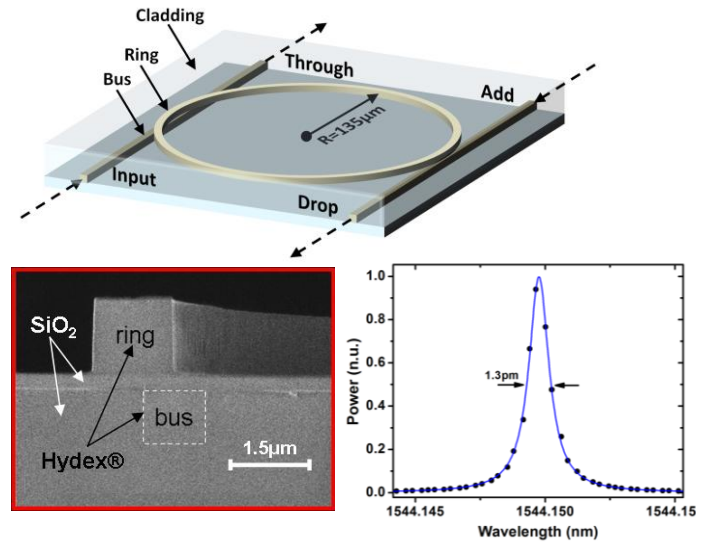

Figure 1. Integrated ring resonator with Q factor of 1.2M as the basis for an integrated multiple waveelength oscillator.

## II. INTEGRATED HYPERPARAMETRIC OSCILLATOR

Figure 1 shows the device structure for the integrated oscillators - a four port micro-ring resonator with radius ≅ 135μm, along with an SEM of the waveguide (1.45μm x 1.5μm). The bus waveguides used to couple light in and out of the resonator have the same cross section and are buried in $SiO_2$. The waveguide core is low loss, high index (n=1.7) doped silica glass with a core-cladding contrast of 17% [13]. The films were deposited by standard chemical vapor deposition (CVD) and device patterning and fabrication were performed by photolithography and reactive ion etching before over-coating with silica glass. Propagation losses were 0.04dB/cm and coupling loss to fiber pigtails of ≅ 1.5dB / facet. Figure 1 also shows the linear transmission spectra of the resonator from the input to the drop port, with a free spectral range (FSR) of 200GHz and a FWHM bandwidth of 1.3pm, corresponding to a Q factor of 1.2 million. The dispersion in these waveguides is extremely low and anomalous [11] over most of the C-band, which is nearly ideal for FWM. The zero dispersion point for TM polarization is ~1560nm in these waveguides, with λ < 1560nm anomalous and λ > 1560nm normal. This provides near ideal conditions to achieve FWM gain.

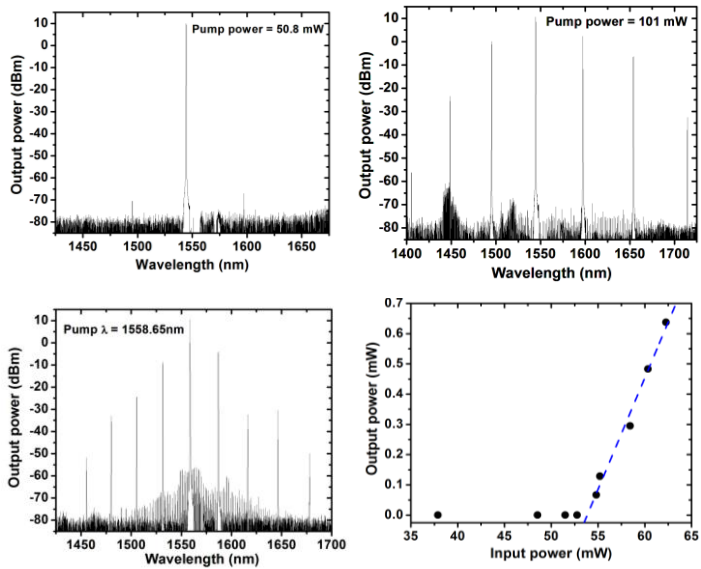

Figure 2. (top) Output spectra of hyperparametric oscillator near threshold (50mW) (top let) and at full pumping power (101mW, top right). Output spectrum when pumping closer to the zero dispersion point (bottom left) and output power of a single line, single port vs pump power.

Figure 2 (top) shows the output spectra for a TM polarized pump beam at 1544.15nm at 50.8mW, just above threshold (top left). Initial lasing occurs at 1596.98nm, 52.83nm away from the pump. As the power is increased above threshold both the wavelength of the mode and the pump power in the cavity are clamped as cascaded FWM takes over, generating further wavelengths with equal spacing. Figure 2 (top right) shows the spectrum at the highest pump power of 101mW (inside the waveguide) showing a wide frequency spacing of almost 53nm. For a pump at 1544.15nm the theoretical MI gain curve peaks at ~1590nm, which agrees very well with the observed initial lasing wavelength. Figure 2 (bottom right) also shows the power of the mode at 1596.98nm exiting one port (drop) versus input pump power, showing a (single line) differential slope efficiency above threshold of 7.4%. The maximum total output power was at 101mW pump at 1544.15nm where we obtained 9mW in all modes out of both ports, with 2.6mW in a single line at 1596.98nm from a single port. The total oscillating mode power of 9mW represents a total conversion efficiency of 9%. The zero dispersion point for TM polarization is ~1560nm in these waveguides, with λ < 1560nm anomalous and λ > 1560nm normal. When pumping at 1565.19 nm (normal dispersion) we observed no oscillation, consistent with parametric gain due to FWM and MI. When pumping near zero dispersion, at 1558.65nm, we observed lasing with a spacing of 28.15nm, agreeing with the expected shift in the MI gain profile (Figure 2 bottom left).

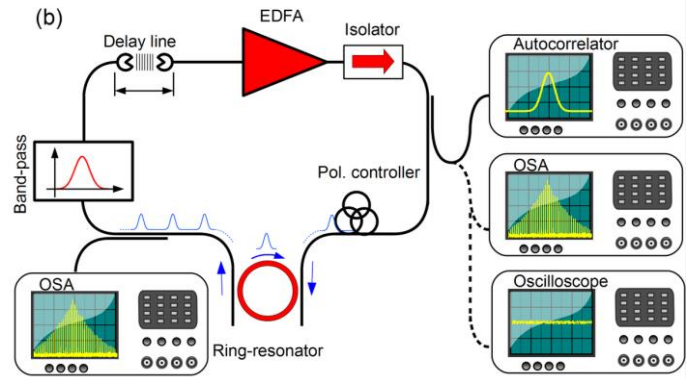

Figure 3. Schematic of soliton modelocked laser based on Q=1.2M ring resonator.

## III. SOLITON LASER

Passively mode-locked lasers have generated the shortest optical pulses to date [20-23]. Self-supporting ultra-short optical pulses can be produced naturally in these lasers by the complex nonlinear interactions between chromatic dispersion, the Kerr nonlinearity and saturable gain. There is considerable interest in achieving both very high and flexible repetition rates [24, 25], at frequencies well beyond the range of active mode-locking where the use of electronics poses inherent limitations. Many different approaches have been proposed to achieve this, ranging from very short laser cavities with large mode frequency spacings (i.e., large Free Spectral Range (FSR) [20, 26-29], where a very high repetition rate is achieved by simply reducing the pulse round-trip time (at the expense of line quality), to schemes where multiple pulses are produced in each round trip [22, 28, 29], applied primarily to fibre lasers. A pioneering approach to the latter was introduced in 1997 by Yoshida et al. [29], where a Fabry Pérot filter inserted in the main cavity suppresses all but a few modes that are periodically spaced, leading to a train of pulses with a controlled repetition rate. This approach was subsequently reinterpreted in terms of the dissipative four-wave-mixing (FWM) paradigm [30-32], and since then, high repetition-rate pulse trains have been demonstrated by adopting variations of this method [31-34]. Although in principle dissipative FWM schemes allow for transform limited pulses to be generated at repetition rates not limited by the main laser cavity, their stability still remains a severe issue - a common problem when multiple pulses circulate in a cavity. This is a consequence of the fact that the gain and nonlinearity required to sustain proper lasing action necessitate the use of very long fibre cavity lengths. This in turn produces smaller cavity mode frequency spacings of (typically) a few megahertz or less, which allows many modes to oscillate within the Fabry Pérot filter bandwidth. Since these modes have random phases, their beating results in severe low frequency noise that produces extremely unstable operation [33].

Figure 3 shows the first mode-locked laser [15] based on a nonlinear monolithic high-Q (quality factor) resonator. This laser achieves extremely stable operation at high repetition rates while maintaining very narrow linewidths, thus leading to a high quality pulsed emission. The key point is that this resonator is not simply used as a filter but acts as the nonlinear element as well, similar in spirit to optically pumped multiple wavelength oscillators based on high Q-factor resonators [35-40]. Due to this twofold central role of the nonlinear filter, this new mode-locking scheme is termed filter-driven four-wave-mixing (FD-FWM). It operates in a way which is in stark contrast to traditional dissipative FWM schemes where the nonlinear interaction occurs in the fibre and is then filtered separately by a linear Fabry Pérot filter.

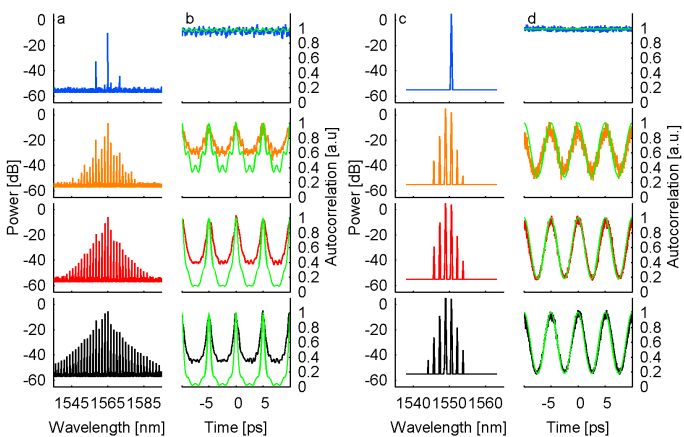

Figure 4. (left) Optical output for long cavity length (unstable) laser and (right0 output of short cavity (stable laser). Green curves are theoretical autocorrelation plots.

The micro-ring resonator is simply embedded in a standard Erbium doped-fibre loop cavity (acting as the gain medium) containing a passband filter with a bandwidth large enough to pass all of the oscillating lines, with the main purpose of controlling the central wavelength. A delay line is employed to control the phase of the main cavity modes with respect to the ring modes.

In order to investigate the dependence of the laser performance on the main cavity length, experiments were performed with two lasers, one based on a short-length erbium-doped fibre amplifier (EDFA) and the other based on a long high-power erbium-ytterbium fibre amplifier (EYDFA), the latter having a similar operating regime to dissipative FWM lasers, where long fibre based cavities are necessary to assure the Kerr nonlinearity required for modelocked laser operation. The two configurations therefore had significantly different main cavity lengths (3 and 33m), i.e. different FSRs (68.5MHz and 6MHz respectively) as well as different saturation powers.

Figure 4 shows the experimental optical spectra of the pulsed output along with the temporal traces obtained by a second order noncollinear autocorrelator for the two systems at four input powers to the ring resonator. The pulses visible in the autocorrelation trains have a temporal duration that decreases noticeably as the input power increases, as expected for a typical passive mode-locking scheme.

From these plots it would appear that the laser based on the high-power EYDFA had superior overall performance since its pulsewidth was considerably shorter. However, this ignores the fact that the autocorrelation measurements average over any long time scale fluctuations of the laser, without requiring a stable pulsed output [40]. The key issue of laser stability or line coherence is better illustrated by a comparison between the experimental autocorrelation traces with the calculated traces (green) for a fully stable and coherent system possessing the optical spectra in Figure 4. While a perfect match is found for the short length EDFA case, the long cavity design shows a considerably higher background, thus clearly distinguishing unstable from stable laser operation.

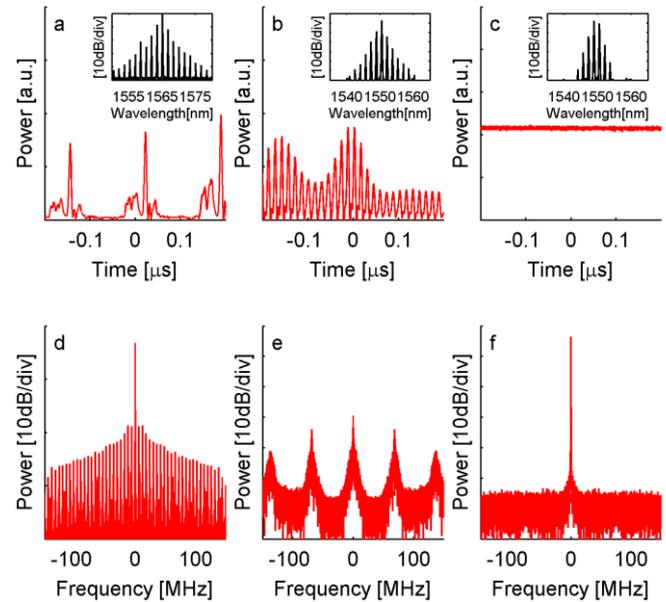

Figure 5. (left) RF plots in time (top) and frequency (bottom) showing unstable behaviour (left) of long cavity, unstable behaviour of short cavity mis-tuned (middle) and stable behaviour of short cavity (right) tuned to align main cavity mode to ring resonator central frequency.

By changing the amplifier driving current we observed a minimum lasing threshold of 500 and 650mA for the unstable and stable designs respectively that corresponded to a gain of approximately 20dB for the both cases. For the stable case, the input power in the ring vs the driving current is plotted in Figure 2(f). To preserve the stable operation the delay line was adapted at the mode-locking threshold. This yielded a slight change in total losses that induced a kink in the plot around a current of 850mA. The maximum input power in the ring was 15.4mW for the stable case, limited by the performance of the EDFA.

To better quantify the pulse-to-pulse amplitude stability of the two lasers we recorded the electrical radio-frequency (RF) spectrum of the envelope signal, collected at the output using a fast photo-detector. Unstable oscillation (in the pulse amplitude) was always observed for the long cavity design for the EYDFA (Figure 5) for both CW and pulsed regimes, due to the presence of a large number of cavity modes oscillating in the ring resonance. In complete contrast to this, the short-cavity configuration for the EDFA could easily be stabilized to give the very clean result of Figure 5 (c,f), by simply adjusting the main cavity length in order to center a single main cavity mode with respect to the ring resonance, thereby completely eliminating any main cavity low-frequency beating. Several stable oscillation conditions were found through tuning the delay by over 2 cm. Note that for the same gain, the optical bandwidth (small black insets in Figure 5 (b,c)) for unstable laser operation was wider (i.e. leading to shorter output pulses) because the instability resulted in a strong amplitude modulation of the optical pulse train in the main cavity, thus increasing the statistical peak power and enhancing the nonlinear interactions.

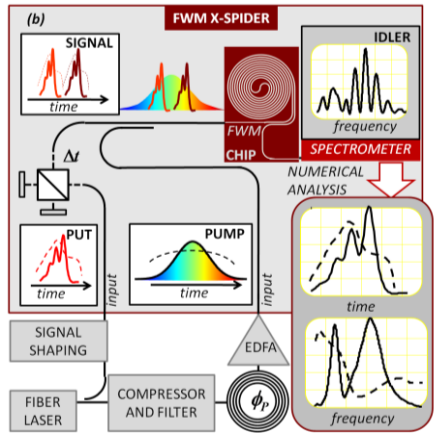

Figure 6. On-chip phase sensitive optical pulse measurement via Spectral Phase Interferometry for Direct Electric-Field Reconstruction (SPIDER) [14].

IV. ULTRAFAST PHASE SENSITIVE PULSE MEASUREMENT

Coherent optical communications has created a need for ultra-fast phase-sensitive measurement at mW powers and on ultrafast pulses. Previous techniques based on integrated platforms include time-lens temporal imaging on a silicon chip [41] and waveguide-based Frequency-Resolved Optical Gating (FROG) [42]. However, the former is phase insensitive while the latter requires long tunable waveguide delay lines which are still an unsolved challenge.

Figure 6 shows a recently reported device [16] capable of both amplitude and phase characterization of ultrafast optical pulses operating via a novel variation of Spectral Phase Interferometry for Direct Electric-Field Reconstruction (SPIDER) [43, 44] based on FWM in the spiral waveguide with the aid of a synchronized incoherently-related clock pulse.

SPIDER approaches are ultrafast and single-shot with a simple and robust phase retrieval procedure. However to date they have used either three-wave mixing (TWM) or linear electro-optic modulation to chirp the pulse, both of which require a 2$^{nd}$ order nonlinearity that is absent in CMOS compatible platforms. Also, SPIDER methods work best with optical pulses shorter than 100fs (small time-bandwidth products (TBP)) and at very high peak powers > 10kW, and hence are not ideally suited to telecommunications. The device in Figure 6 measured pulses with < 100mW peak power, a frequency bandwidth >1THz, and up to 100ps pulsewidths, yielding a record TBP > 100. Figure 7 compares the results from the SPIDER device using both a standard algorithm and a new extended (Fresnel) phase-recovery extraction algorithm [45] designed for pulses with large TBP, with results from SHG-based FROG measurements. As expected, for low TBP pulses (short-pulse regime) the SPIDER device yielded identical results with both algorithms, and agreed with the FROG spectrogram. For large TBP pulses (highly chirped, long pulsewidths) the SPIDER results obtained with the new algorithm agreed very well with the FROG trace, whereas results using the standard phase-recovery algorithm were unable to accurately reproduce the pulse.

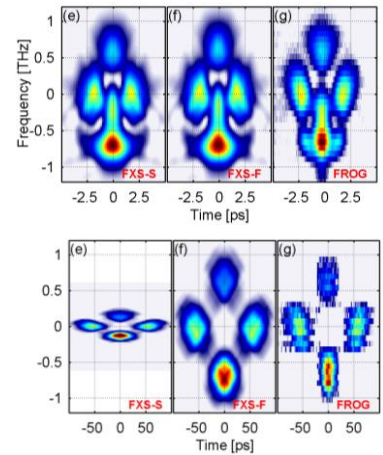

Figure 7. Comparison of SPIDER results with conventional (left) with new (middle) algorithms with FROG results (right) for pulses with low (top) and high (bottom) time bandwidth products (TBP).

V. AMORPHOUS SILICON

The ideal nonlinear optical platform would have all the attributes of silicon-on-insulator but with a FOM >1. Amorphous silicon, of interest as a nonlinear material for some time [46], was recently suggested [47] as a promising alternative to silicon for nonlinear optics, promising a larger FOM than c-Si [47-54]. Recent results have confirmed the possibility of increasing the FOM from around 1 [48,49] to as high as 2 at telecommunication wavelengths [50, 51], enabling high parametric gains of over +26dB over the C-band [50]. Table I surveys the measured nonlinear properties for a-Si, which show a near universal improvement in both the nonlinearity and FOM over c-Si. However, a drawback has been a lack of stability resulting in a dramatic degradation in performance over relatively short timescales [51]. Very recently [52] a-Si nanowires were reported combining a high FOM, nonlinearity and good material stability at telecom wavelengths. Experimental measurements of self-phase modulation and nonlinear transmission revealed both a record high nonlinear FOM of 5 – over 10 x SOI, and a nonlinearity (γ factor), almost 5x silicon. If stable at higher average power levels this material could well represent a practical platform for all-optical devices in the C-band.

## Table I
Amorphous Silicon Nonlinear Properties

| Reference | [52] | [50,51] | [53] | [54] | [48] | [49] |
|---|---|---|---|---|---|---|
| $n_2$ [$10^{-17}$ m$^2$/W] | 2.1 | 1.3 | 4.2 | 0.05 | 0.3 | 7.43 |
| $\gamma$ [W$^{-1}$m$^{-1}$] | 1200 | 770 | 2000 | 35 | N/A | N/A |
| $\beta_{TPA}$ [cm/GW] | 0.25 | 0.392 | 4.1 | 0.08 | 0.2 | 4 |
| FOM | 5 | 2.2 | 0.66 | 0.4 | 0.97 | 1.1 |

This may seem counterintuitive since Kramers - Kronig relations normally imply that increasing the bandgap to decrease nonlinear absorption decreases the nonlinear response. For silicon, however, the real part of the nonlinear susceptibility is largely determined by the direct transitions [55 - 61], while the TPA in the telecom band arises from indirect transitions. For a-Si therefore, it could be hypothesized that an increase in the *indirect* bandgap (reducing TPA) could be accompanied by a decrease in the *direct* bandgap (increasing the Kerr nonlinearity) [62]. Irrespective of these arguments, however, a-Si remains an extremely interesting platform for future all-optical chips.

Finally, a key goal for all-optical chips is to reduce device footprint and operating power, and the dramatic improvement in the FOM of a-Si raises the possibility of using slow-light structures [63 - 67] to allow devices to operate at mW power levels with sub-millimeter lengths.

## VI. CONCLUSION

We demonstrate a wide range of on-chip devices based on a CMOS compatible high-index, doped silica-glass platform. We report a multiple wavelength source based on hyper-parametric oscillation in a microring resonator as well as an ultrahigh repetition rate, stable, modelocked laser, and ultrafast phase sensitive pulse measurement. These devices have significant potential for applications requiring CMOS compatibility for both telecommunications and on-chip WDM optical interconnects for computing.

## ACKNOWLEDGMENT

We acknowledge financial support of the European Union through the Marie Curie program (ALLOPTICS) and the Australian Research Council (ARC).